# Analogues of primeval galaxies two billion years after the Big Bang

Ricardo Amorín[1,2,3,*], Adriano Fontana[1], Enrique Pérez-Montero[4], Marco Castellano[1], Lucia Guaita[1], Andrea Grazian[1], Olivier Le Fèvre[5], Bruno Ribeiro[5], Daniel Schaerer[6,7], Lidia A.M. Tasca[5], Romain Thomas[8], Sandro Bardelli[9], Letizia Cassarà[10,11], Paolo Cassata[8], Andrea Cimatti[12], Thierry Contini[7], Stephane de Barros[6,9], Bianca Garilli[10], Mauro Giavalisco[13], Nimish Hathi[5,14], Anton Koekemoer[14], Vincent Le Brun[5], Brian C. Lemaux[15], Dario Maccagni[10], Laura Pentericci[1], Janine Pforr[5], Margherita Talia[12], Laurence Tresse[5], Eros Vanzella[9], Daniela Vergani[9,16], Giovanni Zamorani[9], Elena Zucca[9], Emiliano Merlin[1]

**Deep observations are revealing a growing number of young galaxies in the first billion year of cosmic time[1]. Compared to typical galaxies at later times, they show more extreme emission-line properties[2], higher star formation rates[3], lower masses[4], and smaller sizes[5]. However, their faintness precludes studies of their chemical abundances and ionization conditions, strongly limiting our understanding of the physics driving early galaxy build-up and metal enrichment. Here we study a rare population of UV-selected, sub-$L^*_{z=3}$ galaxies at redshift $2.4<z<3.5$ that exhibit all the rest-frame properties expected from primeval galaxies. These low-mass, highly-compact systems are rapidly-forming galaxies able to double their stellar mass in only few tens million years. They are characterized by very blue UV spectra with weak absorption features and bright nebular emission lines, which imply hard radiation fields from young hot massive stars[6,7]. Their highly-ionized gas phase has strongly sub-solar carbon and oxygen abundances, with metallicities more than a factor of two lower than that found in typical galaxies of similar mass and star formation rate at $z\leq2.5$[8]. These young galaxies reveal an early and short stage in the assembly of their galactic structures and their chemical evolution, a vigorous phase which is likely to be dominated by the effects of gas-rich mergers, accretion of metal-poor gas and strong outflows.**

[1]INAF-Osservatorio Astronomico di Roma, Via Frascati 33, I-00078, Monte Porzio Catone. Italy
[2]Cavendish Laboratory, University of Cambridge, 19 J.J. Thomson Ave., Cambridge CB30HE, UK
[3]Kavli Institute for Cosmology, University of Cambridge, Madingley Road, Cambridge, CB30HA, UK
[4]Instituto de Astrofísica de Andalucía, CSIC, E-18008 Granada, Spain
[5]Aix Marseille Université, CNRS, LAM (Laboratoire d'Astrophysique de Marseille) UMR 7326, 13388 Marseille, France
[6]Geneva Observatory, University of Geneva, ch. des Maillettes 51, 1290 Versoix, Switzerland
[7]Institut de Recherche en Astrophysique et Planétologie – IRAP, CNRS, Université de Toulouse, UPS-OMP, 14, Avenue E. Belin, 31400 Toulouse, France
[8]Instituto de Física y Astronomía, Facultad de Ciencias, Universidad de Valparaíso, Gran Bretaña 1111, Playa Ancha, Valparaíso, Chile
[9]INAF–Osservatorio Astronomico di Bologna, via Ranzani, 1 - 40127, Bologna, Italy
[10]INAF–IASF, via Bassini 15, 20133 Milano, Italy
[11]Institute for Astronomy, Astrophysics, Space Applications and Remote Sensing, National Observatory of Athens, Penteli, 15236, Athens, Greece
[12]University of Bologna, Department of Physics and Astronomy (DIFA), V.le Berti Pichat, 6/2 – 40127 Bologna, Italy
[13]Astronomy Department, University of Massachusetts, Amherst, MA 01003, USA
[14]Space Telescope Science Institute, 3700 San Martin Drive, Baltimore, MD 21218, USA
[15]Department of Physics, University of California, Davis, One Shields Ave., Davis, CA 95616, USA
[16]INAF–IASF Bologna, via Gobetti 101, 40129 Bologna, Italy
∗ *email*: ra518@mrao.cam.ac.uk

Low-mass star-forming galaxies at the peak epoch of cosmic star formation, at z~1-3, have spectra characterized by faint UV continuum emission exhibiting weak stellar absorption features and strong Lyα emission[9,6]. Unlike more massive $L^*_{z=1-3}$ galaxies, the youngest and most metal-poor galaxies show very blue UV colors and high equivalent widths in nebular emission lines[6,7,10], such as CIII]λλ1906,1909, OIII]λλ1661,1666, HeIIλ1640 and CIVλλ1549,1551, which originate from young HII regions and stellar winds under extreme metallicity and ionization conditions[6,7,10,11]. While galaxies with such unusual spectral properties appear exceedingly rare in spectroscopic surveys[9], increasing evidence suggests that galaxies similar to these are likely to play a preponderant role during the re-ionization era[10,12].

Using the large area and unprecedented sensitivity of the VLT-VIMOS Ultra Deep Survey[13] (VUDS), we search for extremely young, metal-poor galaxy candidates at z>2. We focus in the COSMOS field, for which a wealth of multiwavelength data, including high-resolution HST imaging, is available. From a parent sample of ~1870 VUDS targets at 2.4<z<3.5, we select galaxies with a simultaneous detection (S/N>3) of the CIII], OIII] and Lyα lines in their optical spectra – a selection criterion motivated by the goal of constraining chemical abundances and ionization conditions (Methods). The selected candidates have high Lyα rest-frame equivalent widths (EW>35Å) and most of them also show CIV in emission. Candidates with clear indication of AGN activity according to different diagnostics are excluded (Methods). The final sample consists of 10 galaxies (Supplementary Table 1), which represents ~10% of the galaxies observed by VUDS in the same mass range and < 1% of our total parent sample. Fig. 1 and Fig. 2 illustrate their rest-frame UV spectral and morphological features, respectively.

Chemical abundances and ionization properties of the ionized gas can be derived from electron temperature ($T_e$) measurements or, with less accuracy, using optical strong-line ratios. For galaxies at z=2-4, these indicators are only observable by means of very deep NIR spectroscopy[14,15]. Here, we circumvent the lack of NIR spectra using a novel technique[16,17]. In brief, our method compares a set of observed Lyα, CIII], CIV, and OIII] emission line ratios with those predicted by a detailed grid of photoionization models, which cover an extended range of chemical abundance ratios and ionization conditions (Methods). In the procedure, models are constrained to those values of the carbon-to-oxygen ratio (C/O) and the ionization parameter (U) consistent with the observed ratios. This procedure allows us to derive C/O, U, and metallicity (in units of 12+log(O/H)) in a self-consistent Chi-square scheme, minimizing possible systematics[16]. Using this technique, the selected galaxies are found to have extremely low gas-phase metallicity, low C/O and high ionization parameter (Supplementary Table 2). While log(C/O) is in the range -1.0 to -0.4, with a median log(C/O) = -0.67 well below the solar ratio (log(C/O)$_\odot$ = -0.26)[18], we find a metallicity range of 12+log(O/H) ~ 7.4-7.7 (typical uncertainties ~0.3 dex) that corresponds to ~5%-10% of the solar value (12+log(O/H)$_\odot$=8.7)[18]. These measurements place our sample among the most metal-poor galaxies known at z≥2.4[11].

The estimated carbon and oxygen abundances imply that our galaxies are still chemically young. In evolved, metal-enriched galaxies, carbon is mostly produced by low- and intermediate-mass stars, which increase C/O (and N/O) with increasing metallicity[19,20]. The galaxies in our sample, instead, do not show a correlation between C/O and O/H but rather exhibit a large spread of sub-solar values, a trend that could be explained with models where carbon is essentially produced by massive stars (Fig. 3). Different physical conditions, namely, star formation efficiencies, inflow rates, and variations in the initial mass function may also produce variable levels of C/O at low metallicity[19,20]. The presence of large numbers of hot, massive stars is consistent with the large ionization parameter (median

log($U$)=-2) and the large Lyα EW (up to ~260Å; Supplementary Table 2), which, in the most extreme cases, could suggest a non-standard initial mass function[21] (IMF). The hard-ionizing spectra of massive stellar clusters can also explain the detection of HeII emission in six galaxies, as well as hints of fainter, high-ionization emission lines such as SiIII] λλ1883,1892, SiIV λλ1393,1402, NIII λ1750, and NIV λ1486 in the composite spectrum (Fig. 1). In our sample, HeII is unresolved in the composite spectrum, suggesting relatively narrow lines (FWHM≤1000 km/s). While broad HeII features originate in Wolf-Rayet (WR) stars, narrow HeII emission lines may have a nebular origin[6,22,23,24]. Possible HeII ionizing sources include hot WRs, binaries, shocks from supernovae (SNe) and, especially at very low metallicities, peculiar sources such as a young (Population-III-like) population of very massive or rapidly rotating, metal-poor stars (Z~0.01$Z_\odot$)[23,24]. In either case, chemical enrichment will occur rapidly in these galaxies at the conclusion of the relatively short life span of these stellar populations.

We complement the above measurements with key physical properties such as stellar masses ($M_\star$) and star formation rates (SFR). To this purpose we perform detailed spectral energy distribution (SED) fitting using state-of-the-art multiwavelength photometry and a complete set of stellar and nebular models that account for emission lines and nebular continuum (Methods). This procedure finds our galaxies with low masses $M_\star$~$10^8$-$10^{9.5}$ $M_\odot$, high star formation rates SFR~7-60 $M_\odot$yr$^{-1}$, and low interstellar extinction E(B-V) ≤ 0.15 mag (Supplementary Table 1). Best-fit models suggest very young stellar ages (32 Myr in the median) and unusually high rest-frame EWs for optical emission lines such as Hα and [OIII]λλ4959,5007, which are in most cases larger than ~500Å and produce a significant enhancement of the broad-band fluxes in the NIR (Methods; Supplementary Fig. 1). Consistently, we find very blue UV beta slopes (median β=-2.3) compared to the parent galaxy sample (median β=-1.6, Methods), which suggest very low dust obscuration. For at least 7 out of 10 galaxies, we find SFRs a factor of 2 to 30 times higher than that of normal star-forming galaxies of similar mass at z~3[3] (Methods; Supplementary Fig. 2). Overall, these galaxies have high specific star formation rates (sSFR=SFR/$M_\star$>$10^{-8}$ yr$^{-1}$), which yield a rapid median timescale for the doubling of their stellar content (τ=1/sSFR=30 Myr).

A detailed analysis of spatially resolved HST-ACS images demonstrates that our galaxies are both remarkably compact (Supplementary Table 1; Methods) and diverse morphologically. We find two possible mergers (close pairs with projected separation of <1"), tadpoles, and irregular galaxies, which show one or two bright star forming clumps of less than ~300 parsecs in size (Fig. 2; Methods). The median UV half-light radius of the sample, $r_{50}$=0.5 kpc, is more than a factor of two lower than both the parent galaxy population and typical galaxies of similar stellar mass and redshift (Methods). Such sizes are very similar to those found in young galaxies at z>6[5]. All the galaxies in our sample are spatially resolved in the HST images and show a low surface brightness component that, in some cases (e.g. tadpoles), may suggest a proto-galactic disk. Their total sizes, as measured by the radius accounting for 100% of the light detected in the F814W HST band, range from 0.5 to 2.5 kpc (Supplementary Table 1), which highlights the extreme compactness of these galaxies. Such sizes imply very high SFR surface densities ($\Sigma_{SFR}$=SFR/$2\pi r_{50}^2$≈1-200 $M_\odot$ yr$^{-1}$ kpc$^{-2}$) predicting the need for very high gas surface densities, and thus favor a starburst mode of star formation and strong stellar feedback.

Our young galaxies do not follow the local mass-metallicity relation[25] (MZR), but some of them are otherwise consistent with the low-mass end of the MZR followed by UV-selected galaxies at z>3 (upper panel of Fig. 4), that shows a stronger evolution to low metallicity than observed at lower redshifts[15,26,27]. Even accounting for the intrinsic scatter of the MZR due to different levels of star formation, we find our galaxies a factor of 2-10 offset to lower metallicity from the "fundamental metallicity relation" (FMR[8], bottom panel in Fig. 4). The FMR is a tight relation between stellar mass, gas-phase metallicity and SFR that can be explained by the smooth evolution of galaxies in a quasi-equilibrium state, which is regulated by inflows and outflows over time[8,25]. The observed position of our galaxies in this relation suggests that these galaxies could be driven out of equilibrium by a sudden change in the accretion rate, possibly coupled with strong stellar feedback[27,28]. Detailed simulations of low-mass (~$10^9$ $M_\odot$) galaxies at 2<z<4 show that sudden bursts of star formation are fueled by a massive inflow of metal-poor gas, either from the cosmic web[29] or driven by gas-rich mergers[30] with metal-poor companions, or both. Such a scenario is consistent with the properties observed in our sample, including its morphological diversity. Metal-poor gas accreted from the intergalactic halo may give rise to a bright, compact star-forming clump. During few times the dynamical timescale (of about 20 Myr) the metallicity of this clump will be lower than the surrounding interstellar medium before metal mixing by shear and turbulence driven by disk instabilities restore the galaxies to the equilibrium relation[29] (i.e. the FMR). In similarly short timescales, the collective action of stellar winds and supernovae remnants generate strong metal enriched gas outflows. These outflows would promote the dispersion of metals and the disruption of these small, low mass clumps, in contrast to more massive galaxies hosting long-lived massive clumps[31,32].

The presence of strong outflows combined with intense ionizing radiation and Lyα line emission may result in incomplete coverage of neutral hydrogen (HI), allowing a fraction of the ionizing photons to escape into the intergalactic medium[33]. Despite its low spectral resolution, our composite spectrum provides hints of large velocity outflows of highly ionized gas in our galaxies: We find that low-ionization SiII interstellar absorption lines are significantly blue-shifted with respect to the systemic velocity ($v_{sys}$) determined by CIII], which imply gas moving at velocities of several hundreds of km/s (Methods). In addition, the Lyα peak has only a very small velocity blue-shift of less than 100 km/s with respect to $v_{sys}$, which could be associated to a double-peaked Lyα line with small separation. As we discuss in Methods, this may indicate low HI column densities, which coupled with strong stellar feedback make our galaxies excellent candidates for Lyman continuum emission[11,13,33]. Higher spectral resolution and S/N spectra are required to study in detail gas flows, kinematic properties and their effects in these galaxies.

Our results provide new compelling evidence that searching for faint UV galaxies at z~3 with strong UV metal lines, such as CIII] and OIII], lead us to find very young and metal-poor dwarfs with hard radiation fields and rest-frame properties that are similar to those believed to be common in normal galaxies during the first 500 Myr of cosmic time. These analogues are not pristine but still in a sufficiently young evolutionary stage, before their extreme physical properties will likely change dramatically on timescales of few million years. Thus, they may provide unique insight on the earliest phases of galaxy formation and the role of low-mass star-forming galaxies to the reionization of the Universe.

## Methods

**Observations, sample selection, and line measurements:**

The *VIMOS Ultra Deep Survey*[13] is a deep spectroscopic legacy survey of ~10,000 galaxies carried out using VIMOS at ESO-VLT. This survey is aimed at providing a complete census of the SF galaxy population at 2≤z≤7, covering ~1 square degree in three fields: COSMOS, ECDFS, and VVDS-2h. The VIMOS spectra consist of 14h integrations in the LRBLUE and LRRED grism settings, covering a combined wavelength range 3650Å<λ<9350Å, with a spectral resolution R~230. Data reduction, redshift measurement, and assessment of the reliability flags are described in detail in the survey and data presentation papers[13,34].

For this work, we identify a representative sample of 10 galaxies showing strong UV emission lines in the portion of the COSMOS field observed by VUDS. Our selection criteria are motivated by the aim of looking for extremely metal-poor galaxy candidates at z~3 from the observed optical spectroscopy[35,6,36,37,38]. This requires measuring rest-frame UV emission line ratios tracing carbon, oxygen and hydrogen abundances[39,40,41], and the ionization parameter[17] (as discussed below). Thus, our goal is to find galaxies showing the following emission lines: CIII] λλ1907,1909, OIII] λλ1661,1666, CIV λλ1549,1551, and Lyα.

The selected galaxies are first extracted from a parent sample of 1870 galaxies in the redshift range 2.4≤z≤3.5, which allows for the simultaneous observation of the lines in the central, high S/N portion of the VUDS spectra. From this parent sample, we select 870 galaxies with very reliable spectroscopic redshift (≥95% probability to be correct)[13]. We excluded objects with relevant emission lines affected by strong sky-subtraction residuals and galaxies identified as AGN in X-ray surveys or with clear signs of activity from emission line several emission line diagnostics (see below). Thus, 10 galaxies are selected for their simultaneous detection (S/N>3) of Lyα, CIII] λλ1907,1909 and OIII] λλ1661,1666 emission lines. It is worth noticing that these carbon and oxygen inter-combination doublets (hereinafter CIII] and OIII]) are unresolved due to the limited spectral resolution. Similarly, we find that 8 out of 10 of these galaxies are also detected with S/N≥3 in CIV λλ1549,1551 (hereinafter CIV).

The long integration time of ~14h per target in VUDS allow us to detect the continuum at ~8000Å (~2000Å at z=3) with S/N~5 for galaxies down to $i_{AB}$~25 and emission lines with fluxes down to F~1.5x10$^{-18}$ erg/s/cm$^2$/Å (S/N~5)[13]. This permits the identification of other very faint emission lines, such as the HeII λ1640 and NIII λ1750, and a few absorption features in individual spectra (Supplementary Fig. 4). However, faint low-ionization absorption lines, such as SiII λ1260 and SiII λ1526 (see below), are only clearly detected in the composite spectrum shown in Fig. 1. The composite spectrum is generated by stacking the 10 individual spectra, after deriving the systemic redshift from the observed CIII] line as the mean centroid of the Gaussian fit to each line. Then, all science and noise spectra are re-binned to a dispersion of 1.4 Å per pixel, which corresponds to the VIMOS pixel scale divided by (1+$z_{med}$), where $z_{med}$ is the median systemic redshift. Finally, we average combine the interpolated rest-frame science spectra and generate a composite noise spectrum by summing in quadrature the science spectra in flux units and dividing this quantity by the number of spectra to combine.

Down to the flux (EW) limits for the VUDS survey, our selected sample represents about <1% of the total parent sample and ~10% of parent sample galaxies in the same stellar mass range. This, and the strong relation existing between the CIII] and OIII] intensities and metallicity and ionization parameter for a given C/O[42], imply that we are likely detecting the most metal-poor, high-ionization galaxies of our parent sample.

For this work, we use emission-line integrated fluxes that are measured manually on a one by one basis. The Lyα rest-frame EWs listed in Table 2 agree with those published in Cassata *et al.* (2015)[43]. Flux measurements are done using the IRAF task *splot* and adopting an integration of the line profile after linear subtraction of the continuum, which is detected (S/N>2) in all cases. Instead of using the noise spectrum computed by the data reduction pipeline, uncertainties in the line measurements are computed from the dispersion of values provided by multiple measurements adopting different possible band-passes (free of lines and strong residuals from sky subtraction) for the local continuum determination, which is fitted using a second order polynomial. We note that the adopted uncertainties are typically larger than those obtained from the average noise spectrum produced by the data reduction pipeline. To compute line ratios, we first apply a reddening correction to the observed fluxes. We use the Calzetti *et al.* (2000)[44] extinction curve and assume that $E(B-V)_{gas}=E(B-V)_\star$, where $E(B-V)_\star$ is obtained from the SED fitting (see below).

## Ionization source: Star formation vs. AGN

By construction, the presence of high ionization emission lines in our galaxy sample is indicative of a hard-ionizing spectrum. To constrain the dominant ionization source of the galaxies we perform a number of tests. We look for the presence of AGN activity using (i) detection in the deepest X-ray surveys available and (ii) emission line diagnostics based on the comparison of observed UV emission line ratios (e.g. CIV/CIII, CIV/HeII) with those predicted by detailed photoionization models. Additionally, for the only source included in the CANDELS[45] footprint, VUDS-5100998761, we look for flux variability between the original COSMOS HST F814W band images[46] and the latest CANDELS images[47] (timescale of ~ 10 years). We do not find any significant change (>3σ) in its ACS-F814W magnitudes. From our first criterion, we disfavor bright AGNs as ionizing sources because of the lack of X-ray counts at the same HST coordinates in the Chandra images of the COSMOS-legacy catalogue[48,49] with an effective exposure time of ~160ks (limiting luminosity $L_x>10^{43}$ erg/s). The lack of detection is confirmed using the stacked image of the ten galaxies in the soft band, which has a mean count rate of about 1.4 times the statistical error. Moreover, none of these galaxies show broad emission lines (FWHM>1200 km/s), which exclude the possibility of a broad-line AGN. Although lower luminosity or obscured narrow-line AGNs cannot be entirely ruled out, we are not able to demonstrate their presence with current data.

Our second criterion relies on the recent UV emission line diagnostics of Feltre *et al.* (2016)[50] and Gutkin *et al.* (2016)[51] that are based on photoionization models of active and non-active galaxies, respectively. As shown in Supplementary Fig. 3, we find our ten galaxies and the average composite spectrum (Fig. 1) in the region essentially populated by non-active models, i.e. star-forming galaxies, in four different diagnostics involving CIV, CIII], OIII] and HeII. As reference, the combination of emission line ratios (CIV/CIII] =0.6, CIV/HeII=1.2, CIII]/HeII=2.2, OIII]/HeII=1.2, and Lyα/CIV=28.6) measured in the composite spectrum appear different from the typical ratios observed in AGNs[52]. Similar conclusions are found when comparing our data against other recent photoionization models by Jaskot & Ravindranath (2016)[42], which also include the contribution of shock ionization.

Additional reasons, such as their extremely blue UV-to-IR SED and their low mass, low dust, and low gas-phase metallicity, point to a stellar ionizing source as the likely dominant ionization source, although some contribution from non-stellar ionizing sources (e.g. shocks) cannot be ruled out, especially in those objects with the highest CIII] equivalent widths. Future observations, in particular NIR spectroscopy, combined with our current data and detailed photoionization models will provide additional constrains on the dominant ionizing sources.

## Systemic redshift and velocity shifts: Probing outflows

We investigate line velocity shifts that may suggest the presence of gas outflows by using our composite spectrum (Fig. 1). A proper determination of the systemic velocity and the relative shifts of interstellar (IS) low ionization absorption lines tracing outflows are not possible in individual spectra due to the limited S/N of the lines. However, using stacking for our ten sources we can improve the detection of both emission and absorption features and derive a rough estimate of the systemic velocity and shifts. As explained before, since stellar photospheric absorption lines are not detected we adopt the centroid of CIII] for the systemic redshift determination when doing stacking. Since CIII] is a doublet, we assumed a ratio for the two components of CIII] λ1907/[CIII] λ1909 = 1.5, which appear appropriate for the range of electron densities ($n_e \leq 100$ cm$^{-3}$) and electron temperatures ($T_e \geq 15000$ K) we expect to find in our galaxies[53]. At the VUDS resolution (R<300), a large variation of this ratio only affects our measurement in a relatively small velocity uncertainty of a few tens of km/s.

Then we derive the velocity shift $\Delta v_{IS}=v_{IS}-v_{CIII]}$ for the low ionization IS absorption lines SiII λ1260 and SiII λ1526, with detection at the ~3-5σ level (EW=1.6±0.3 and 1.3±0.4, respectively). For these two lines, we find blue shifts of 3.0±0.4 Å and 3.5±0.9 Å, which translate into outflow velocities of -715±95 km/s and -687±176 km/s for SiII λ1260 and SiII λ1526, respectively, and an average outflow velocity of $\Delta v_{IS} \sim -700\pm136$ km/s. Although hints for other absorption lines such as CII λ1334 or SiIV+OIV λ1397 are observed, they are poorly resolved and have S/N ratios that appear insufficient to provide additional significance to the above results. Our results, however, are consistent with those obtained from the stacking of a larger population of CIII] emitters in VUDS (Guaita *et al.*, in prep.). A more detailed analysis of outflows velocities for our sample will require additional high S/N data with higher spectral resolution.

## Implications for the escape of ionizing photons

The inferred outflow velocity $\Delta v_{IS}$ for our sample of galaxies is higher than the typical values for Lyman-break galaxies (LBGs) of ⟨Δv⟩ = -150 km/s reported by Shapley *et al.* (2003)[9]. However, our $\Delta v_{IS}$ is consistent with their mean difference between Lyα emission and IS absorption lines, Δ(em-abs)~Δ($v_{Ly\alpha}$ - $v_{IS}$)= 650 km/s, of LBGs with Lyα emission. Indeed, we find that $v_{CIII]}$ and $v_{Ly\alpha}$ differ in only ~50±90 km/s, which at the limited resolution of the spectra (R<300) is an indication that Lyα is close to the systemic velocity, in clear contrast with more normal galaxies at similar redshift (~445 km/s, Steidel et al. (2010)[54]). This may also indicate the presence of a substantial blue-shifted component in a double-peaked Lyα emission with small separation, which at the resolution of VUDS could be observed as a small blueward shift. This may have interesting implications for the escape of ionizing photons into the intergalactic medium. These small shifts appear indicative of strong (i.e. high EW) Lyα emitting galaxies with compact star-forming regions, high ionization and low-metallicity[55,56,57,58,59]. They also appear associated to strong, high-velocity outflows[60] of highly ionized gas and low HI covering fraction[61,62], which may allow Lyα photons to emerge without being substantially scattered or absorbed. The above conditions also appear to promote the escape of Lyman continuum (LyC) photons, as predicted by models[63,64] and shown by larger resolution spectra of a few extreme emission line galaxies with escaping LyC emission at low[65,66] and high redshift[11,12].

## SED fitting: Stellar masses, star formation rates and ages

We estimate physical properties by fitting the observed multiwavelength photometry with a set of Bruzual & Charlot (2003, hereafter BC03)[67] synthetic models through a $\chi^2$ minimization routine called ZPHOT[68,69,70] (Supplementary Fig. 1). We use the latest photometric catalogue available in COSMOS[71], which includes deeper UltraVista DR2 and Spitzer-IRAC photometry, adding some technical improvements with respect to their predecessors, e.g. in source extraction and de-blending. The IRAC bands are particularly important to constrain the fits in the rest-frame NIR part of the SED and give robustness to our stellar mass and age determination. We find all galaxies in the sample to be detected in the IRAC 3.6 and 4.5 micron bands, while in the 5.8 and 8 micron bands galaxies are undetected, excepting for VUDS-510838687 and VUDS-51011421970, which are also detected at 5.8 microns. For the latter, however, the IRAC fluxes may suffer of contamination due to the halo of a nearby bright star. Thus, its physical parameters should be considered with caution and, in particular, the stellar mass and ages should be considered as upper limits. Also, it is worth mentioning that at least one of the galaxies appears as a merging system (VUDS-510838687) and their SED properties correspond to the system. The photometric apertures considered (2 arcsec) exceed the size of the system in the HST F814W image.

For each object, we set the redshift to the spectroscopic value. The stellar templates are produced using four possible stellar metallicity values ($Z/Z_\odot$=0.02, 0.2, 0.4, 1). We adopt a Chabrier et al. (2003)[72] IMF, a Calzetti et al. (2000)[44] extinction law, and a range of physical parameters: 0≤E(B-V)≤1.1 and ages ≥0.01 Gyr (defined as the onset of the star formation). The star formation history (SFH) has been parameterized by (i) an exponentially declining law or "tau" model (proportional to exp(-t/τ)) with timescale τ=0.1,0.3,0.6,1.0,2.0,5.0,15.0 Gyr, (ii) an exponentially rising or "inverted tau" model (proportional to exp(t/τ)) law with the same timescales, and (iii) a constant SFH.

In addition to the stellar template we have included the contribution from nebular emission following Schaerer & de Barros (2009)[73], which is constrained by the number of hydrogen-ionizing photons in the stellar SED (Schaerer & Vacca 1998)[74] assuming an escape fraction of stellar LyC ionizing photons $f_{esc}$=0. The ionizing radiation is converted in nebular continuum considering free-free, free-bound and H two-photon continuum emission, assuming an electron temperature $T_e$=10$^4$ K, an electron density $N_e$=100 cm$^{-3}$, and abundance of Helium relative to Hydrogen of 10%. Hydrogen lines are included considering case B recombination, while the relative line intensities of He and metals as a function of metallicity are taken from Anders & Fritze-v.Alvensleben (2003)[75].

Stellar masses and ages from the SED fitting are presented in Table 1. All the ten sample galaxies are well fitted by the models, with reduced $X^2$ close to unity. Several consistency checks are performed to test the robustness of the fits against e.g. the specific code used for the SED fitting (we also use LePhare[76] and GOSSIP+[77]), the initial set of model parameters (e.g. age, metallicity, E(B-V) ranges), the shape of the assumed SFH, and the addition of a second component accounting for a maximally old stellar population. Overall differences in the relevant physical parameters resulting among these tests yield values that are typically consistent within the quoted uncertainties.

**UV slopes and SFR**

We derive the UV slope β using the four optical bands covering the rest-frame UV portion between 1200Å and 2000Å following Castellano et al. (2014)[69] and Hathi et al. (2016)[78]. For the parent sample, we used as a reference β=-1.6, as derived by Hathi et al. (2016)[78]. The UV beta slopes for our sample of galaxies are included in Table 1. We adopt the IRX relations of Castellano et al. (2014)[69] for dust attenuation corrections, which implies a dust-free UV slope as steep as β =-2.67. We find the dust attenuations computed from β and the E(B-V) obtained from the SED fitting in agreement within the uncertainties.

Following Talia et al. (2015)[79], we use the dust attenuation to derive the dust-corrected UV luminosity. We adopt the standard calibration of Kennicutt (1998)[80] to convert UV luminosities into SFR, after diving its normalization by a factor of 1.7 to scale down from a Salpeter to a Chabrier IMF. The UV- and SED-based SFRs agree within the 68% confidence level. The latter values are used to compute the median values presented in the main text and the bottom panel of Fig. 4. Both SFR estimates are presented in Table 1.

**Chemical abundances and ionization conditions**

In order to derive oxygen and carbon abundances using UV indicators, we use an updated version of the code HII-CHI-MISTRY (HCM[16]), which is presented in a companion paper (Pérez-Montero & Amorín, 2017)[17]. Hereinafter, we refer to this version as HCM-UV. In brief, using the observed emission lines HCM-UV computes a set of line indices, which depend on metallicity and ionization. These indices are then compared with predictions from a large grid of CLOUDY v13 photoionization models[81] covering a wide range of possible physical conditions, O/H and C/O abundances, and ionization conditions. The HCM-UV code has the ability of using UV line indices in addition (or alternatively) to the standard optical line indices adopted in the optical version. This allows HCM-UV to compute first C/O and then derive metallicity and the ionization parameter using only UV metal line tracers. In HCM-UV, models are constrained by empirical relations found between C/O, O/H and the ionization parameter ($U$) for a wide range of metallicity (from few % solar to super-solar) and $U$ values, thus minimizing possible systematics[16,17]. The code uses the *Pyneb v*0.9.3 software[82] for computing density, temperature, ionic abundances and ionization correction factors. Thus, C/O and then O/H can be derived together with $U$ as the $X^2$-weighted mean of these quantities in the models, with uncertainties provided by the corresponding $X^2$-weighted standard deviation. This method yields larger but more realistic uncertainties than that obtained by the direct method from the propagation of flux errors into their standard expressions. In general, results obtained with HCM-UV for a sample of local and high redshift emission line galaxies with UV and optical line measurements[17], are typically consistent within ~0.1 dex with $T_e$-based abundances, such as those derived through the direct method or strong-line calibrations using objects with reliable measurements of the electron temperature[16]. When a determination of the electron temperature is not possible (e.g. due to the lack of optical emission lines such as Hβ and [OIII] λ5007), uncertainties are larger (~0.2 dex on average) but still consistent with the scale of O/H and C/O abundances provided by the direct method.

For this work, four UV nebular lines are available: Lyα, CIII], CIV, and OIII]. Then, HCM uses three line indices defined as C34=log((CIII]+CIV)/Lyα), C3O3=log(CIII]/OIII]), and C3C4=log(CIII]/CIV). C34 has a similar dependence with log(O/H) as the widely used optical index R23=log([OII]+[OIII]/Hβ), with two possible values of metallicity below and above a knee located around metallicity of about 20% solar. To break possible degeneracies, we use C3O3 and C3C4, which account for the relation between metallicity and both C/O and the ionization parameter, respectively[17,83]. Additionally, the code has the ability of using optical lines such as Hβ, [OII] λλ3727,3729, and [OIII]λλ4959,5007. For instance, the index RO3=log(OIII] λ1666/[OIII] λ5007) is a proxy for $T_e$[84], while using Hβ instead of Lyα in C34 for the hydrogen abundance can reduce uncertainties due to the complex radiative transfer of Lyα. Several tests have been performed on the few galaxies available in the literature with a complete set of UV plus optical emission lines and for which a $T_e$-based metallicity and C/O is available[17]. Overall, results are in good agreement within a typical error of ~0.3 dex, although the agreement tends to be better for galaxies with high Lyα EW. It is worth noticing that errors in the Lyα flux due to e.g. scattering or strong attenuation, translate to an error in metallicity by a similar factor.

## Morphology and sizes

Morphological parameters, i.e. total radii $r_{T100}$, effective radii $r_e$, and axis ratio $q$, are taken from Ribeiro et al. (2016)[85]. They were computed from the ACS-F814W HST mosaics[46] (limiting AB magnitude of 27.2 for a 5σ point-source detection) using both a parametric fitting, in the case of $r_e$ and $q$, with the code GALFIT[86] and a non-parametric, model-independent method in the case of $r_{T100}$. The latter considers the circularized radius that encloses 100% of the measured flux above a certain surface brightness threshold and appears as a useful, complementary information to the effective radii computed by GALFIT (see Ribeiro et al. (2016)[85] for further details). This method allows us to demonstrate that, although they are extremely compact, all the sample galaxies are spatially resolved in the HST-ACS images. One of the most compact objects in the sample, VUDS-5100998761, is in the CANDELS footprint. This allows us to compute its morphological parameters using the CANDELS mosaics, which include the WFC-F125W band of the NIR[47]. Our results confirm its compact morphology, with optical and NIR sizes in good agreement within the uncertainties.

## Acknowledgments

We acknowledge the two anonymous referees for very constructive and helpful reports. We also thank Veronica Sommariva for her contribution to the initial steps of this work. This work is supported by funding from the European Research Council Advanced Grant ERC-2010-AdG-268107-EARLY and by INAF Grants PRIN 2010, PRIN 2012 and PICS 2013.This work is based on data products made available at the CESAM data center, Laboratoire d'Astrophysique de Marseille, France. This research leading to these results has received funding from the European Union Seventh Framework Programme ASTRODEEP (FP7 2007/2013) under grant agreement n° 312725. R.A. acknowledges support from the ERC Advanced Grant 695671 "QUENCH". EPM acknowledges support from Spanish MICINN grants AYA2010-21887-C04-01 and AYA2013-47742-C4-1-P. Based on data obtained with the European Southern Observatory Very Large Telescope, Paranal, Chile, under Large Program 185.A-0791. The complete spectroscopic dataset of this program is available through the ESO archive: http://archive.eso.org/cms.html. A description of the survey and a database containing catalogues and data products is available at: http://cesam.lam.fr/vuds/. Results on the chemical abundances have been obtained using our own publicly available code HCm, which is available at: http://www.iaa.es/~epm/HII-CHI-mistry.html

## References

1. Bowens, R. J. et al., UV Luminosity Functions at Redshifts z~4 to z~10: 10,000 Galaxies from HST Legacy Fields. *Astrophys. J.* **803**, 34-49 (2015)

2. Smit, R. et al., Evidence for Ubiquitous High-equivalent-width Nebular Emission in z~7 Galaxies: Toward a Clean Measurement of the Specific Star-formation Rate Using a Sample of Bright, Magnified Galaxies. *Astrophys. J.* **784**, 58, (2014)

3. Tasca, L. A. M. et al., The evolving star formation rate: M. relation and sSFR since z ≃ 5 from the VUDS spectroscopic survey. *Astron. Astrophys.* **581A**, 54 (2015)

4. Grazian, A. et al., The galaxy stellar mass function at 3.5 ≤z ≤ 7.5 in the CANDELS/UDS, GOODS-South, and HUDF fields. *Astron. Astrophys.*, **575**, A96 (2015)

5. Shibuya, T. *et al.*, Morphologies of ~190,000 Galaxies at z = 0-10 Revealed with HST Legacy Data. I. Size Evolution. *Astrophys. J. Suppl.*, **219**, 15 (2015)

6. Erb, D. *et al.*, Physical Conditions in a Young, Unreddened, Low-metallicity Galaxy at High Redshift. *Astrophys. J.*, **719**, 1168-1190 (2010)

7. Stark, D. *et al.*, Ultraviolet emission lines in young low-mass galaxies at z ≃ 2: physical properties and implications for studies at z > 7. *Mon. Not. R. Astron. Soc.*, **445**, 3200 (2014)

8. Mannucci, F. *et al.* A fundamental relation between mass, star formation rate and metallicity in local and high-redshift galaxies. *Mon. Not. R. Astron. Soc.*, **408**, 2115-2127 (2010)

9. Shapley, A. *et al.*, Rest-Frame Ultraviolet Spectra of z~3 Lyman Break Galaxies. *Astrophys. J.*, **588**, 65-89 (2003)

10. Vanzella, E. *et al.*, High-resolution Spectroscopy of a Young, Low-metallicity Optically Thin L = 0.02L* Star-forming Galaxy at z = 3.12. *Astrophys. J. Letts.*, **821**, L27 (2016)

11. de Barros, S. *et al.*, An extreme [O III] emitter at z = 3.2: a low metallicity Lyman continuum source. *Astron. Astrophys.*, **585**, A51 (2016)

12. Vanzella, E. *et al.*, Hubble imaging of the ionizing radiation from a star-forming galaxy at z=3.2 with fesc>50%. *Astrophys. J.*, **825**, 41 (2016)

13. Le Fèvre, O., *et al.*, The VIMOS Ultra-Deep Survey: ~10 000 galaxies with spectroscopic redshifts to study galaxy assembly at early epochs 2 < z <~ 6. *Astron. Astrophys.*, **576**, A79 (2015)

14. Steidel, C. C. *et al.*, Strong Nebular Line Ratios in the Spectra of z ~ 2-3 Star Forming Galaxies: First Results from KBSS-MOSFIRE. *Astrophys. J.*, **795**, 165, (2014)

15. Onodera, M. *et al.*, ISM Excitation and Metallicity of Star-forming Galaxies at z ≃ 3.3 from Near-IR Spectroscopy. *Astrophys. J.*, **822**, 42 (2016)

16. Perez-Montero, E., Deriving model-based $T_e$-consistent chemical abundances in ionized gaseous nebulae. *Mon. Not. R. Astron. Soc.*, **441**, 2663-2675 (2014)

17. Pérez-Montero, E. & Amorín, R., Using photo-ionisation models to derive carbon and oxygen abundances in the rest UV. *Mon. Not. R. Astron. Soc.*, accepted (2017)

18. Asplund, M., *et al.*, The Chemical Composition of the Sun. *Ann. Rev. Astron. Astrophys.*, **47**, 481-522 (2009)

19. Mattsson, L. *et al.*, The origin of carbon: Low-mass stars and an evolving, initially top-heavy IMF?. *Astron. Astrophys.*, **515**, A68 (2010)

20. Berg, D. A., *et al.*, Carbon and Oxygen Abundances in Low Metallicity Dwarf Galaxies. *Astrophys. J.*, **827**, 126 (2016)

21. Malhotra, S. & Rhoads, J. E., Large equivalent width Lyα line emission at z=4.5: Young galaxies in a young universe?. *Astrophys. J. Letts.*, **565**, 71-74 (2002)

22. Steidel, C. *et al.*, Reconciling the stellar and nebular spectra of high-redshift galaxies. *Astrophys. J.*, **826**, 159 (2016)

23. Cassata, P. *et al.*, HeII emitters in the VIMOS VLT Deep Survey: Population III star formation or peculiar stellar populations in galaxies. *Astron. Astrophys.*, **556**, A68, (2013)

24. Kehrig, C. *et al.*, The extended HeII λ4686-emitting region in IZw 18 unveiled: Clues for peculiar ionizing sources. *Astrophys. J. Letts.*, **801**, 28-34, (2015)

25. Andrews, B.H., & Martini, P., The Mass-Metallicity Relation with the Direct Method on Stacked Spectra of SDSS Galaxies. *Astrophys. J.*, **765**, 140 (2013)

26. Maiolino, R., *et al.*, AMAZE. I. The evolution of the mass-metallicity relation at z > 3. *Astron. Astrophys.*, **488**, 463 (2008)

27. Troncoso, P. *et al.*, Metallicity evolution, metallicity gradients, and gas fractions at z~3.4. *Astron. Astrophys.*, **563**, A58 (2014)

28. Sánchez Almeida, J. *et al.*, Localized Starbursts in Dwarf Galaxies Produced by the Impact of Low-metallicity Cosmic Gas Clouds. *Astrophys. J. Letts.*, **810**, L15 (2016)

29. Ceverino, D. *et al.*, Gas inflow and metallicity drops in star-forming galaxies. *Mon. Not. R. Astron. Soc.*, **457**, 2605-2612 (2016)

30. Bekki, K., Formation of blue compact dwarf galaxies from merging and interacting gas-rich dwarfs. *Mon. Not. R. Astron. Soc. Letts.*, **388**, L10-14 (2008)

31. Bournaud, F., *et al.*, The Long Lives of Giant Clumps and the Birth of Outflows in Gas-rich Galaxies at High Redshift. *Astrophys. J.*, **780**, 57 (2014)

32. Zanella, A., *et al.*, An extremely young massive clump forming by gravitational collapse in a primordial Galaxy. *Nature*, **521**, 54-56 (2015)

33. Erb, D., Feedback in low-mass galaxies in the early Universe. *Nature*, **523**, 169-176 (2014)

34. Tasca, L.A.M., *et al.*, The VIMOS Ultra Deep Survey First Data Release: spectra and spectroscopic redshifts of 698 objects up to z~6 in CANDELS. *Astron. Astrophys.* submitted (eprint arXiv:1602.01842)

35. Fosbury, R. A. *et al.*, Massive Star Formation in a Gravitationally Lensed H II Galaxy at z=3.357. *Astrophys. J.*, **596**, 797-809, (2003)

36. Christensen, L. *et al.*, Gravitationally lensed galaxies at 2 < z < 3.5: direct abundance measurements of Lyα emitters. *Mon. Not. R. Astron. Soc.*, **427**, 1973-1982 (2012)

37. James, B. L., *et al.*, Testing metallicity indicators at z~1.4 with the gravitationally lensed galaxy CASSOWARY 20. *Mon. Not. R. Astron. Soc.*, **440**, 1794-1809 (2014)

38. Bayliss, M. B., *et al.*, The Physical Conditions, Metallicity and Metal Abundance Ratios in a Highly Magnified Galaxy at z = 3.6252. *Astrophys. J.*, **790**, 144 (2014)

39. Garnett, D., *et al.*, High Carbon in I Zwicky 18: New Results from Hubble Space Telescope Spectroscopy. *Astrophys. J.*, **481**, 174-178 (1997)

40. Garnett, D., *et al.*, Carbon in Spiral Galaxies from Hubble Space Telescope Spectroscopy. *Astrophys. J.*, **513**, 168-179 (1999)

41. Pettini, M., *et al.*, C, N, O abundances in the most metal-poor damped Lyman alpha systems. *Mon. Not. R. Astron. Soc.*, **385**, 2011-2024 (2008)

42. Jaskot, A. E. & Ravindranath, S., Photoionization Models for the Semi-Forbidden CIII] 1909 Emission in Star-Forming Galaxies. *Astrophys. J.*, submitted (2016) (eprint ArXiv:1610.03778)

43. Cassata, P., *et al.*, The VIMOS Ultra-Deep Survey (VUDS): fast increase in the fraction of strong Lyman-α emitters from z = 2 to z = 6. *Astron. Astrophys.*, **573**, A24 (2015)

44. Calzetti, D., *et al.*, The Dust Content and Opacity of Actively Star-forming Galaxies. *Astrophys. J.*, **533**, 682-695 (2000)

45. Grogin, N. A. *et al.*, CANDELS: The Cosmic Assembly Near-infrared Deep Extragalactic Legacy Survey. *Astrophys. J. Suppl.*, **197**, 35 (2011)

46. Koekemoer, A. M. *et al.*, The COSMOS Survey: Hubble Space Telescope Advanced Camera for Surveys Observations and Data Processing. *Astrophys. J. Suppl.*, **172**, 196-202 (2007)

47. Koekemoer, A. M. *et al.*, CANDELS: The Cosmic Assembly Near-infrared Deep Extragalactic Legacy Survey—The Hubble Space Telescope Observations, Imaging Data Products, and Mosaics. *Astrophys. J. Suppl.*, **197**, 36 (2011)

48. Civano, F., The Chandra Cosmos Legacy Survey: Overview and Point Source Catalog. *Astrophys. J.*, **819**, 62 (2016)

49. Marchesi, S., *et al.*, The Chandra COSMOS Legacy survey: optical/IR identifications. *Astrophys. J.*, **817**, 34 (2016)

50. Feltre, A., Charlot, S., Gutkin, J., Nuclear activity versus star formation: emission-line diagnostics at ultraviolet and optical wavelengths. *Mon. Not. R. Astron. Soc.*, **456**, 3354-3374 (2016)

51. Gutkin, J., Charlot, S., Bruzual, G., Modelling the nebular emission from primeval to present-day star-forming galaxies. *Mon. Not. R. Astron. Soc.*, **462**, 1757-1774, (2016)

52. Hainline, K.N., *et al.*, The Rest-frame Ultraviolet Spectra of UV-selected Active Galactic Nuclei at z ~ 2-3. *Astrophys. J.*, **733**, 31 (2011)

53. Keenan, F. P.; Feibelman, W. A.; Berrington, K. A., Improved calculations for the CIII 1907, 1909 and Si III 1883, 1892 electron density sensitive emission-line ratios, and a comparison with IUE observations. *Astrophys. J.*, **389**, 443-446

54. Steidel, C. *et al.*, The Structure and Kinematics of the Circumgalactic Medium from Far-ultraviolet Spectra of z ~= 2-3 Galaxies. *Astrophys. J.*, **717**, 289-322 (2010)

55. Hashimoto, T., *et al.* Gas Motion Study of Lyα Emitters at z ~ 2 Using FUV and Optical Spectral Lines, *Astrophys. J.*, **765**, 70 (2013)

56. Shibuya, T., *et al.*, What is the Physical Origin of Strong Lyα Emission? II. Gas Kinematics and Distribution of Lyα Emitters. *Astrophys. J.*, **788**, 48 (2014)

57. Erb, D. *et al.*, The Lyα Properties of Faint Galaxies at z ~ 2-3 with Systemic Redshifts and Velocity Dispersions from Keck-MOSFIRE. *Astrophys. J.*, **795**, 33 (2014)

58. Trainor, R., *et al.*, The Spectroscopic Properties of Lyα-Emitters at z ˜2.7: Escaping Gas and Photons from Faint Galaxies. *Astrophys. J.*, **809**, 89 (2015)

59. Bradac, M., et al., ALMA [CII] detection of a redshift 7 lensed galaxy behind RXJ1347.1-1145. *Astrophys. J. Letts.*, submitted (eprint arXiv:1610.02099)

60. Heckman, *et al.* Extreme Feedback and the Epoch of Reionization: Clues in the Local Universe. *Astrophys. J.*, **730**, 5 (2011)

61. Jaskot, A. E. & Oey, M. S., Linking Lyα and Low-ionization Transitions at Low Optical Depth. *Astrophys. J. Letts.*, **791**, L19 (2014)

62. Henry, A., *et al.*, Lyα Emission from Green Peas: The Role of Circumgalactic Gas Density, Covering, and Kinematics. *Astrophys. J.*, **809**, 19 (2015)

63. Verhamme, A., *et al.*, Using Lyman-α to detect galaxies that leak Lyman continuum. *Astron. Astrophys.*, **578**, A7 (2015)

64. Dijkstra, M.; Gronke, M.; Venkatesan, A. The Lyα-LyC Connection: Evidence for an Enhanced Contribution of UV-faint Galaxies to Cosmic Reionization. *Astrophys. J.*, **828**, 71 (2016)

65. Borthakur, S. *et al.* A local clue to the reionization of the universe. *Science*, **346**, 216-219 (2014)

66. Verhamme, A., *et al.*, Lyman-alpha spectral properties of five newly discovered Lyman continuum emitters. *Astron. Astrophys.*, accepted (2016) (eprint ArXiv: 1609.03477)

67. Bruzual, G. & Charlot, S., Stellar population synthesis at the resolution of 2003. *Mon. Not. R. Astron. Soc.*, **344**, 1000-1028 (2003)

68. Fontana, A., *et al.*, A European Southern Observatory Very Large Telescope Survey of Near-Infrared (Z <= 25) Selected Galaxies at Redshifts 4.5 < z < 6: Constraining the Cosmic Star Formation Rate near the Reionization Epoch. *Astrophys. J.*, **587**, 544-550 (2003)

69. Castellano, M., *et al.*, Constraints on the star-formation rate of z ~ 3 LBGs with measured metallicity in the CANDELS GOODS-South field. *Astron. Astrophys.*, **566**, A19 (2014)

70. Santini, P., *et al.*, Stellar Masses from the CANDELS Survey: The GOODS-South and UDS Fields. *Astrophys. J.*, **801**, 97 (2015)

71. Laigle, C., *et al.*, The COSMOS2015 Catalog: Exploring the 1<z<6 Universe with half a million galaxies. *Astrophys. J. Suppl. S.*, **224**, 24 (2016)

72. Chabrier, G., Galactic Stellar and Substellar Initial Mass Function. *Public. Astron. Soc. Pac.*, **115**, 763-795 (2003)

73. Schaerer, D. & de Barros, S., The impact of nebular emission on the ages of z≈ 6 galaxies. *Astron. Astrophys.*, **502**, 423-426 (2009)

74. Schaerer, D. & Vacca, W.D., New Models for Wolf-Rayet and O Star Populations in Young Starbursts. *Astrophys. J.*, **497**, 618-644 (1998)

75. Anders, P., & Fritze-v. Alvensleben, U., Spectral and photometric evolution of young stellar populations: The impact of gaseous emission at various metallicities. *Astron. Astrophys.*, **401**, 1063-1070 (2003)

76. Ilbert O., *et al.*, Cosmos Photometric Redshifts with 30-Bands for 2-deg$^2$. *Astrophys. J.*, **690**, 1236-1249 (2009)

77. Thomas, R., *et al.*, The VIMOS Ultra-Deep Survey (VUDS): IGM transmission towards galaxies with 2.5<z<5.5 and the colour selection of high redshift galaxies, *Astron. Astrophys.* accepted (eprint ArXiv: 1411.5692T)

78. Hathi, N. *et al.*, The VIMOS Ultra Deep Survey: Lyα emission and stellar populations of star-forming galaxies at 2<z<2.5. *Astron. Astrophys.*, **588**, A26 (2016)

79. Talia, M., *et al.*, The star formation rate cookbook at 1 < z < 3: Extinction-corrected relations for UV and [OII]λ3727 luminosities. *Astron. Astrophys.*, **582**, A80 (2015)

80. Kennicutt, R.C Jr., Star Formation in Galaxies Along the Hubble Sequence. *Ann. Rev. Astron. Astrophys.*, **36**, 189 (1998)

81. Ferland, G.J., *et al.*, The 2013 Release of Cloudy. *Rev. Mex. Astron. Astrof.*, **49**, 137-163 (2013)

82. Luridiana, V., Morrisett, C. & Shaw, R.A., PyNeb: a new software for the analysis of emission lines. *Planetary Nebulae: An Eye to the Future, Proceedings of the International Astronomical Union, IAU Symposium*, **283**, 422-423 (2012)

83. Garnett, D., *et al.*, The evolution of C/O in dwarf galaxies from Hubble Space Telescope FOS observations. *Astrophys. J.*, **443**, 64-76 (1995)

84. Villar-Martín, M., Cerviño, M. & González-Delgado, R., Nebular and stellar properties of a metal-poor HII galaxy at z= 3.36. *Mon. Not. R. Astron. Soc.*, **355**, 1132-1142 (2004)

85. Ribeiro, B. *et al.*, Size evolution of star-forming galaxies with 2 < z < 4.5 in the VIMOS Ultra-Deep Survey. *Astron. Astrophys.*, **593**, A22 (2016)

86. Peng, C.Y., *et al.*, Detailed Structural Decomposition of Galaxy Images. *Astron J.*, **124**, 266-293 (2002).

87. Roberts-Borsani, G. W., *et al.*, z ≳ 7 Galaxies with Red Spitzer/IRAC [3.6]-[4.5] Colors in the Full CANDELS Data Set: The Brightest-Known Galaxies at z ~ 7-9 and a Probable Spectroscopic Confirmation at z = 7.48. *Astrophys. J.,* **823**, 143 (2016)

88. Huang, K-H, *et al.*, Spitzer UltRa Faint SUrvey Program (SURFS UP). II. IRAC-detected Lyman-Break Galaxies at 6 ≲ z ≲ 10 behind Strong-lensing Clusters. *Astrophys. J.,* **817**, 11 (2016)

89. Amorín, R., *et al.*, Evidence of Very Low Metallicity and High Ionization State in a Strongly Lensed, Star-forming Dwarf Galaxy at z= 3.417. *Astrophys. J. Letts.*, **788**, L4 (2014)

90. Amorín, R., *et al.*, Extreme emission-line galaxies out to z ~ 1 in zCOSMOS. I. Sample and characterization of global properties. *Astron. Astrophys.*, **578**, A105 (2015)

91. Maseda, M. V., *et al.*, The nature of emission line galaxies at z=1-2: Kinematics and metallicities from near-infrared spectroscopy. *Astrophys. J.*, **791**, 17 (2014)

92. Schreiber, C. *et al.*, The Herschel view of the dominant mode of galaxy growth from z=4 to the present day. *Astron. Astrophys.*, **575**, A74, (2015)

## Figures

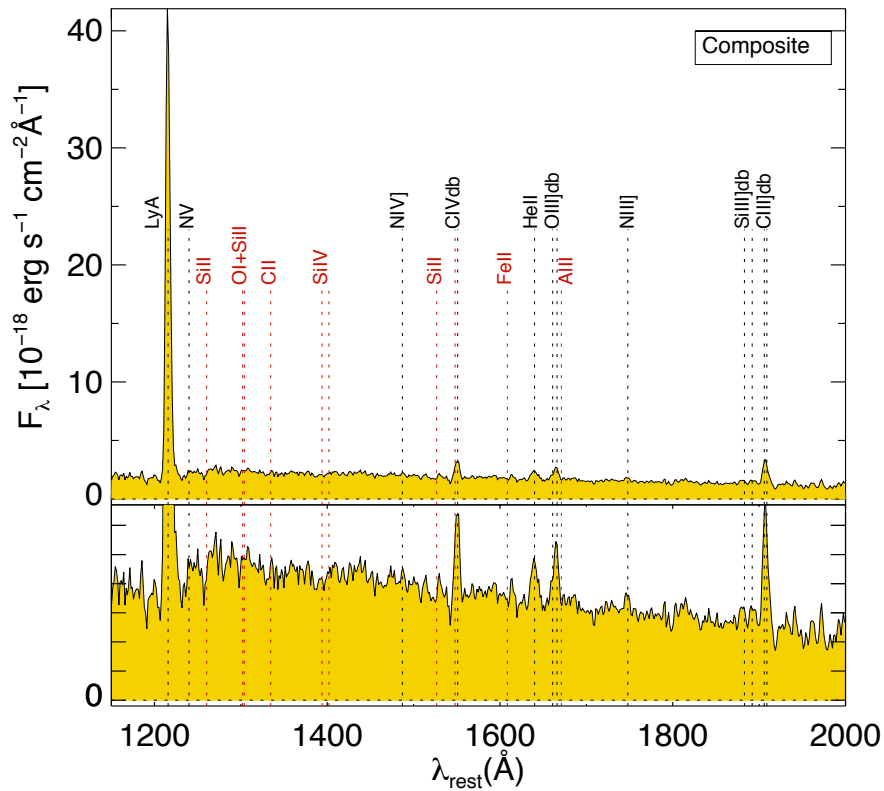

**Fig. 1 | Composite spectrum of the ten sample galaxies.** Lyman-alpha and the three UV metal lines in emission (CIV1548,1551, OIII]1661,1666, CIII]1907,1909) used for the abundance analysis are labeled in black, along with other relevant emission lines. The rest wavelengths of relevant UV absorption lines are labeled in red. The composite spectrum is produced through stacking of individual spectra (Methods), which are presented in Supplementary Information.

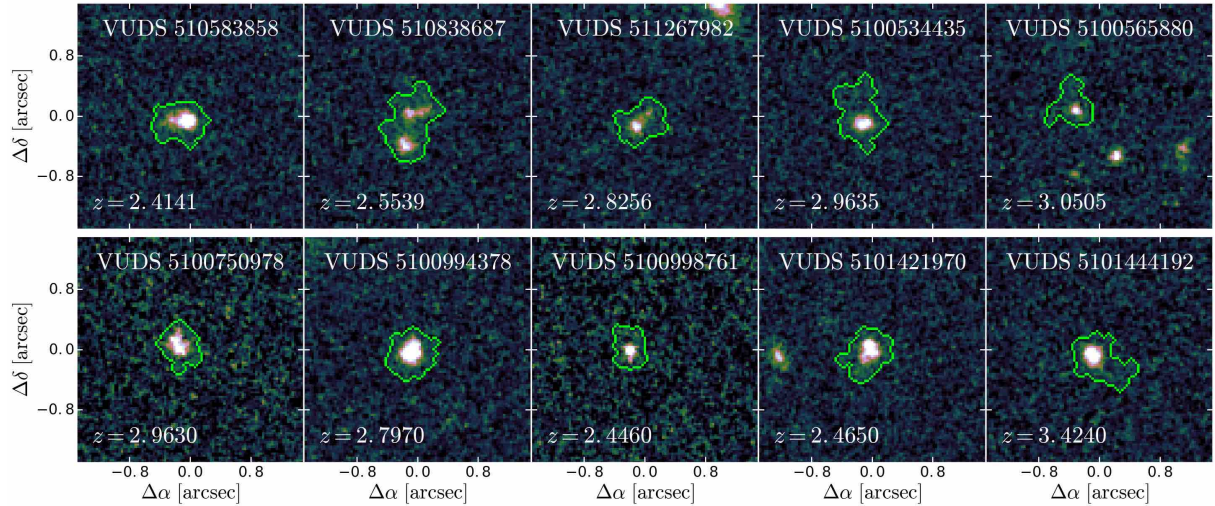

**Fig. 2 | The UV morphologies.** HST-ACS F814W (FWHM~0.03") postage stamps of the ten galaxies of the sample. At the redshift of the galaxies, F814W corresponds approximately to the rest-frame near-ultraviolet (NUV) band. The green contour show, for each sample galaxy, the isophote containing 100% of the light included in the detection images (Methods). All the galaxies are extremely compact, showing one or two bright star forming clumps in a very low-surface brightness irregular component (limiting AB magnitude of 27.2; Methods). From the top-left to bottom-right corners we identify tadpole shapes (#1, 2, 3, 4, 9, 10), close pairs (#2, 5), and single clumps (#6, 7, 8), which illustrate the morphological diversity of the sample.

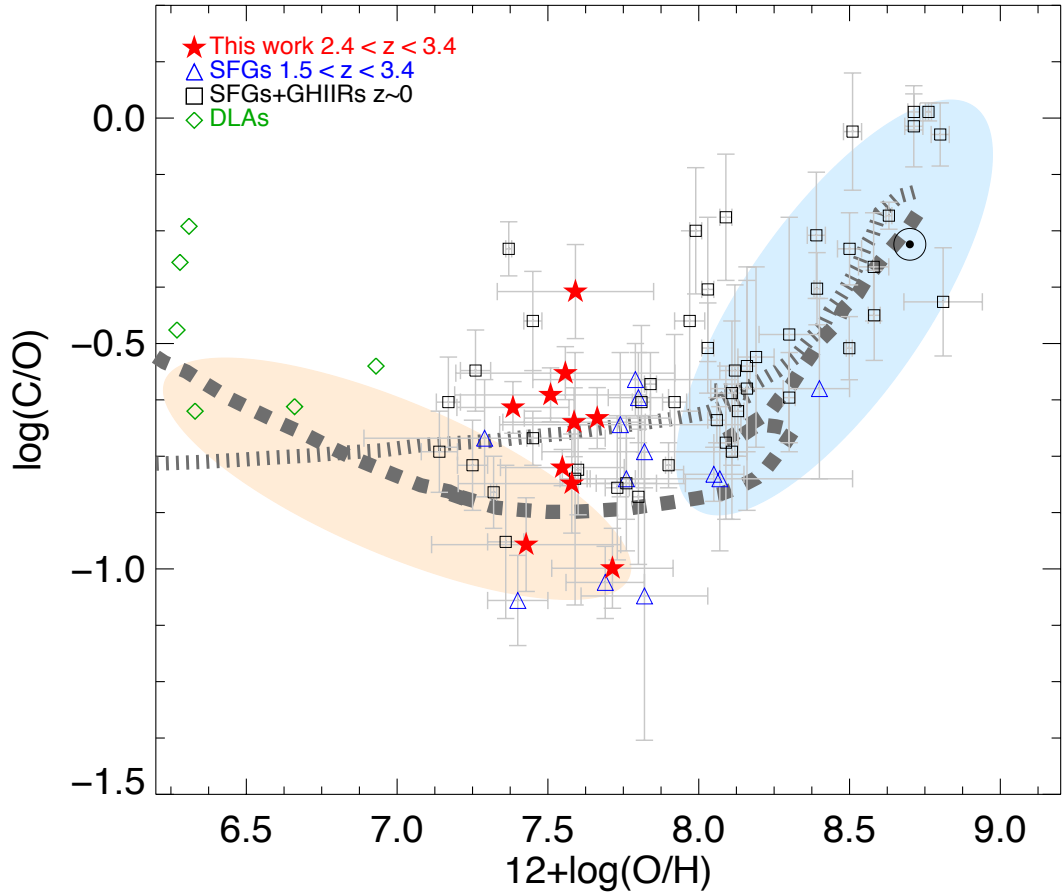

**Fig. 3 | The C/O vs. O/H relation.** The orange and blue shadowed areas correspond to the trend followed by metal-poor and metal-rich stars of the Milky Way's halo and disk, respectively[19]. Dotted and dashed thick curves show models from Mattsson *et al.* (2012)[19], where carbon is mostly produced by high mass stars (model B1) and low-to-intermediate mass stars with an evolving IMF (model E1), respectively. For comparison, we include abundances compiled in previous studies[17,19,20] from the literature for both strongly lensed and non-lensed galaxies at 1<z<3, damped Lyman-α (DLA) systems, and local star-forming galaxies and giant extragalactic HII regions (GHIIRs). In all cases metallicities were obtained through measurements of $T_e$, while C/O values were estimated using the same index C3O3 used in the present work. Error bars account for observational (emission line ratios) and methodological (statistical) 1σ uncertainties.

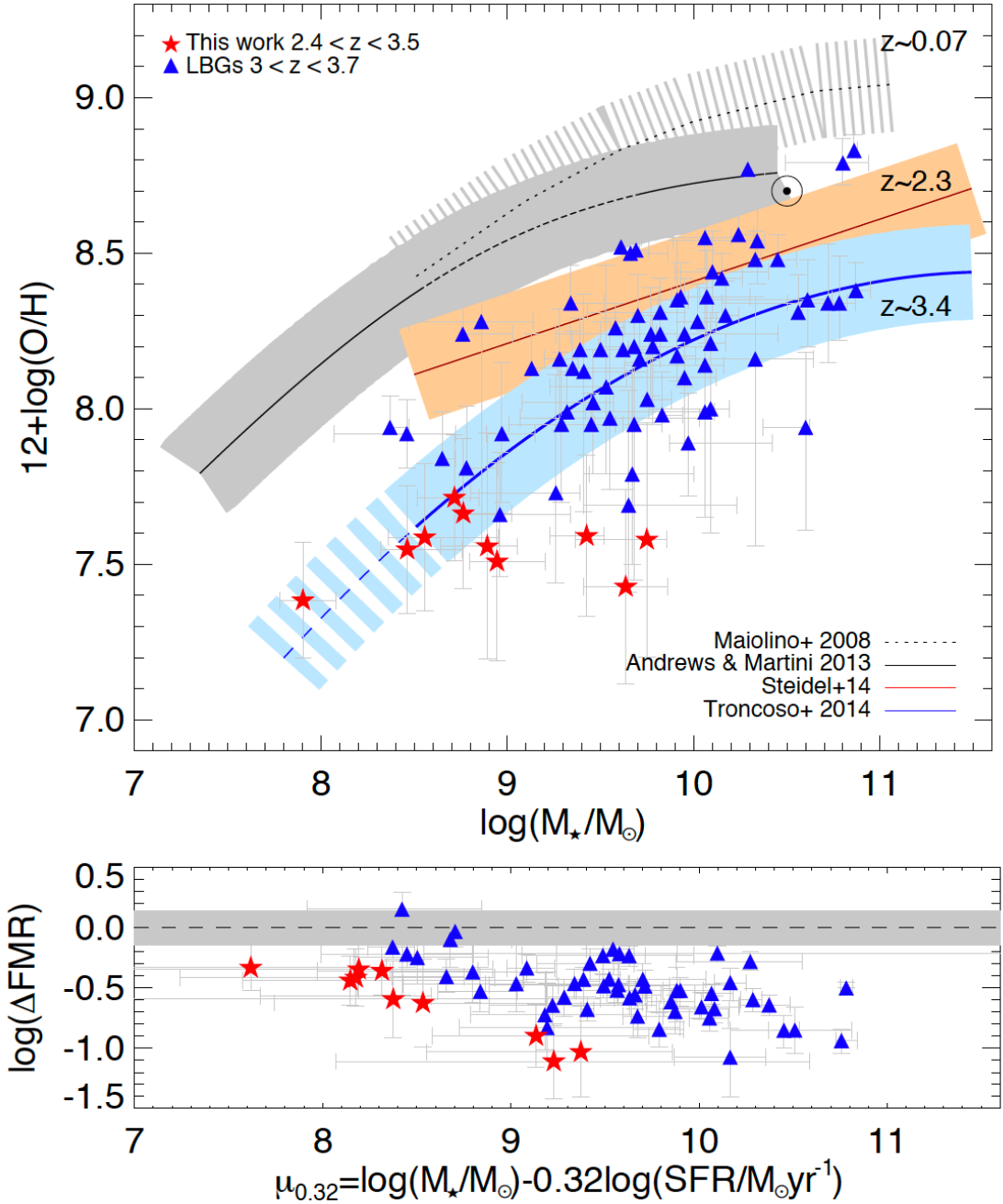

**Fig. 4 | The relation between stellar mass, gas-phase metallicity, and SFR.** The position of our sample of galaxies (red stars) and other UV-selected star-forming galaxies at z>3 from the literature[15,27] (blue triangles) is shown along with the average mass-metallicity relation (MZR) at various redshifts[14,25,26,27] (Upper panel) and the inferred difference in metallicity from the fundamental metallicity relation, FMR[8,25] (i.e., $\Delta$FMR=[O/H]$_{obs}$-[O/H]$_{FMR}$; Bottom panel). The latter is shown as a function of $\mu_{32}$. Shadowed regions indicate typical 1$\sigma$ uncertainties for the various MZR and the FMR. Literature data gathered for the comparison sample include galaxies with metallicities obtained through $T_e$-consistent strong-line methods, along with stellar masses and SFRs inferred through SED fitting using models and procedures consistent with those used in the present work. In both cases, error bars account for observational (emission line ratios) and methodological (i.e., statistical) 1$\sigma$ uncertainties.

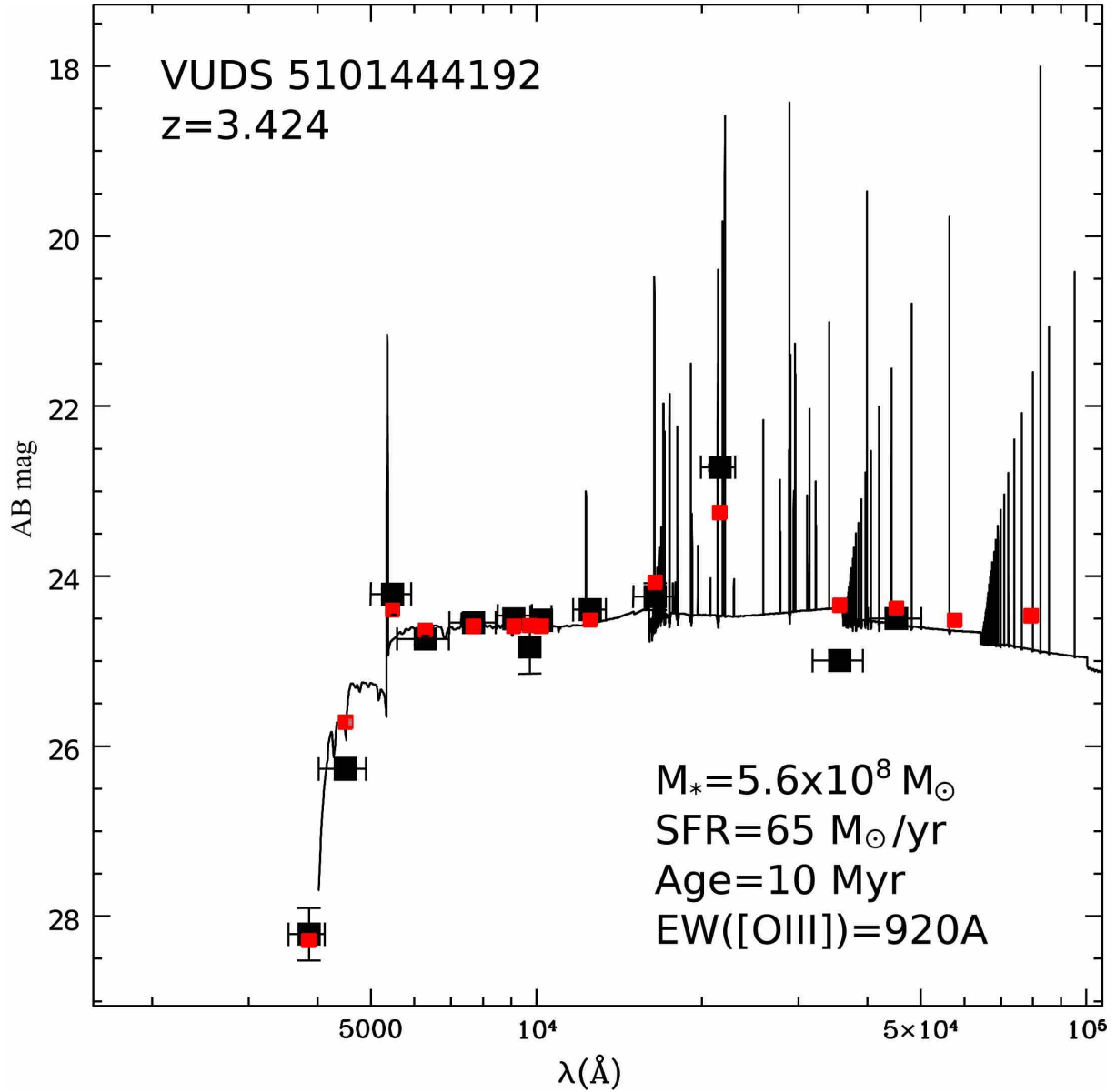

**Supplementary Fig. 1 | Photometric SED fit of the most distant galaxy of the sample.** Best-fit model points to the observed spectral energy distribution spanning rest-frame UV to near-IR are shown by red and black squares, respectively. Error bars account for 1σ uncertainties in the broad-band photometry. Labels indicate ID number, spectroscopic redshift, and the main output physical parameters (from top to bottom: stellar mass, star formation rate, luminosity-weighted stellar age, and rest-frame [OIII]5007 equivalent width) from the SED fitting. Note the remarkable jump of ~2 magnitudes produced by the contribution of unusually high equivalent width emission lines (EW([OIII]+Hβ)~1500Å) to the observed H band at the redshift of the galaxy. While such spectral features appear relatively common at higher redshift (z>6)[2,87,88], at z<3.5 they typically define the rare class of extreme emission-line galaxies[89,90,91].

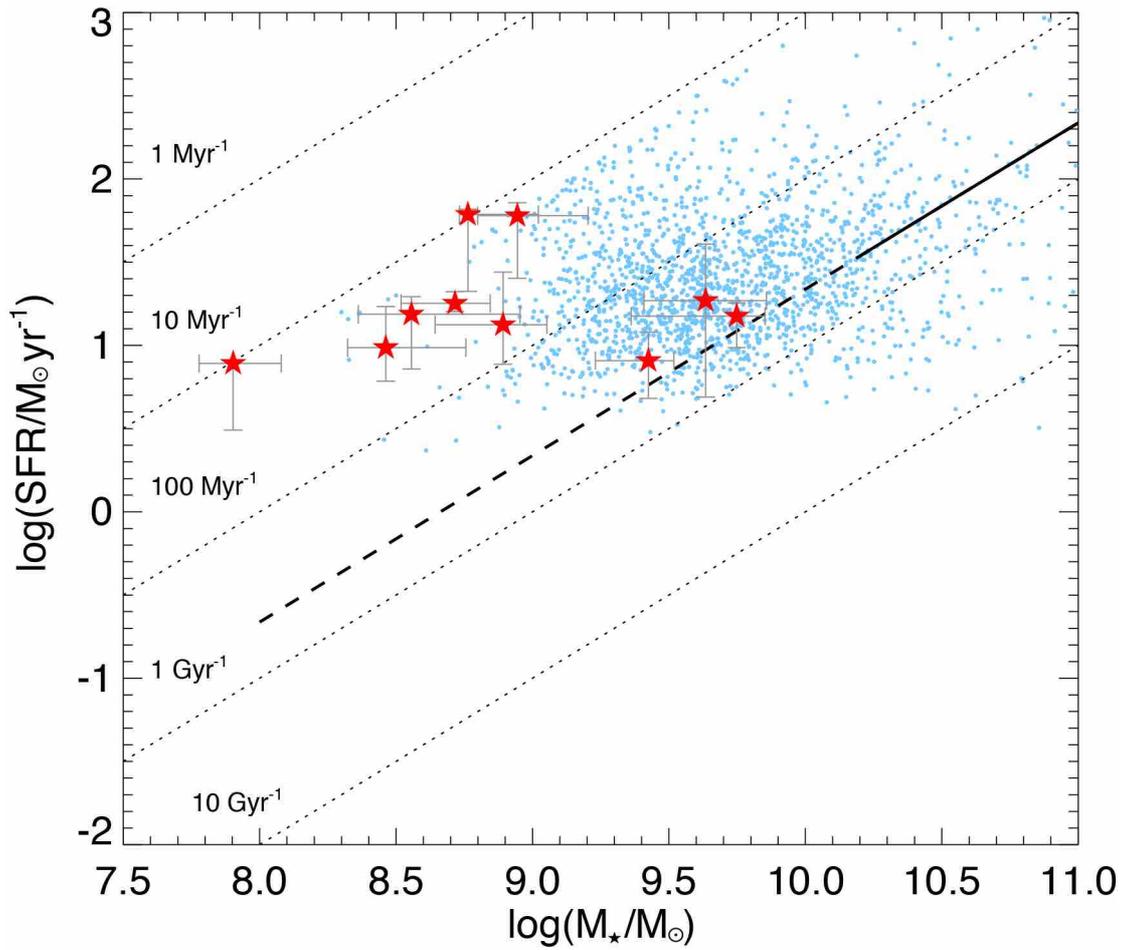

**Supplementary Fig. 2 | The SFR vs. stellar mass plane.** We show the position of our galaxy sample (red stars) and the parent sample used in this work (blue dots), which includes all galaxies in the portion of the COSMOS field observed by VUDS in the same redshift range, 2.4<z<3.5. Both SFRs and stellar masses for this figure have been derived using the same SED fitting technique (Methods). Error bars account for 1σ uncertainties. For comparison, the black solid line indicates the star-forming main sequence of galaxies at z~2.5-3.5 and its extrapolation to low stellar masses (dashed line) from Schreiber *et al.* (2015)[92].

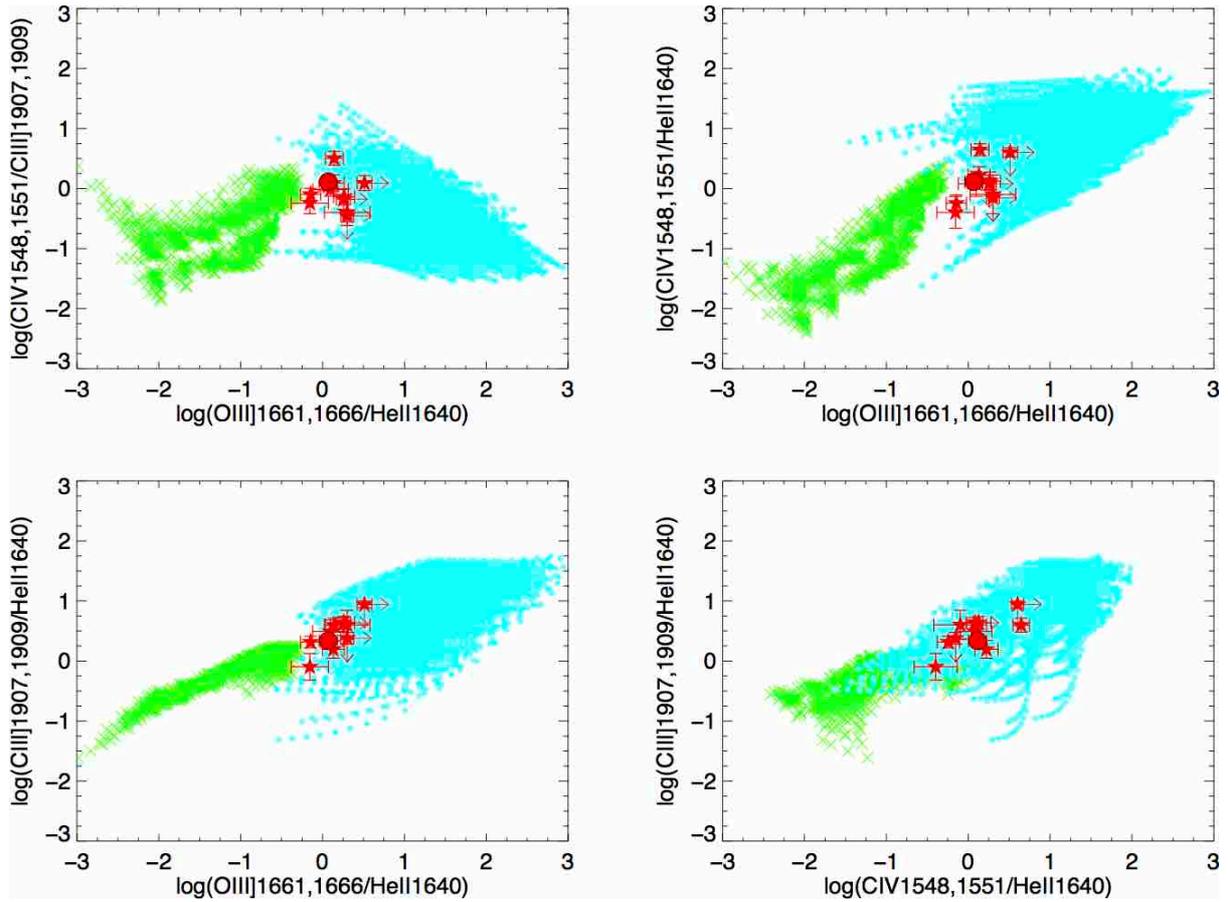

**Supplementary Figure 3 | UV emission line-ratio diagnostic diagrams**. The distributions of narrow-lined AGN (green crosses) and star-forming (turquoise dots) models of Feltre *et al.* (2016)[50] and Gutkin *et al.* (2016)[51], respectively, span full ranges in the input parameters as described in these papers, excepting for the interstellar gas metallicities, which are selected from Z=0.0002 (12+log(O/H)~7) to Z=0.02 (12+log(O/H)~9). Red stars show values for individual galaxies, while the line ratios computed from the composite spectrum are indicated as a red circle. Error bars account for 1$\sigma$ uncertainties in the emission line ratios.

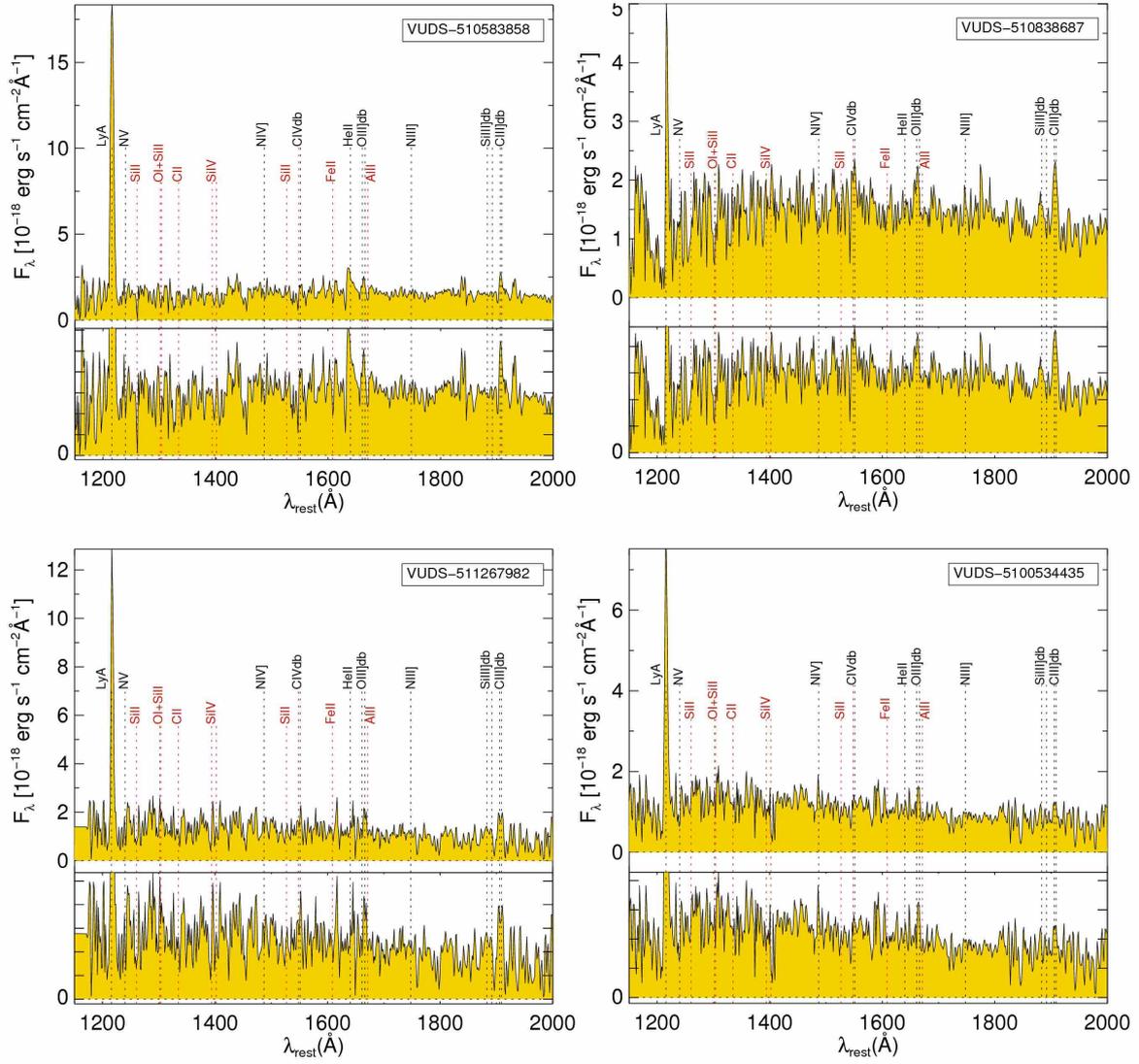

**Supplementary Figure 4 | Individual VUDS spectra.** Labels indicate line features as in Figure 1 and identification numbers as in Supplementary Table 1.

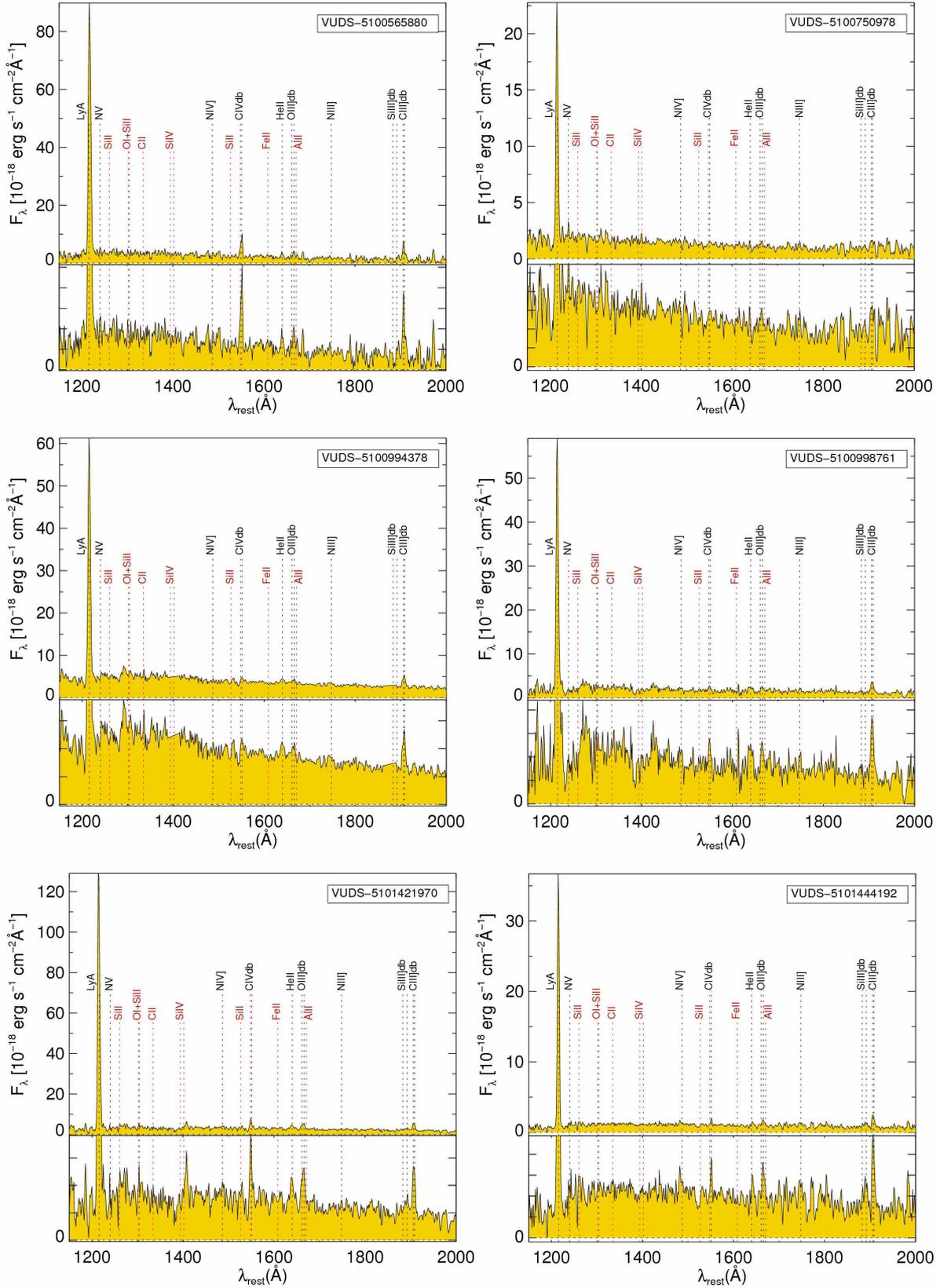

**Supplementary Figure 4 | Individual VUDS spectra** *(Continued).*

| VUDS ID (1) | z (2) | i [mag] (3) | $R_e$ [kpc] (4) | $R_{T100}$ [kpc] (5) | $M_\star$ [$10^9 M_\odot$] (6) | SFR$_{SED}$ [$M_\odot$yr$^{-1}$] (7) | SFR$_{UV}$ [$M_\odot$yr$^{-1}$] (8) | $E(B-V)_\star$ mag (9) | $\beta_{UV}$ (10) | Age Myr (11) |
|---|---|---|---|---|---|---|---|---|---|---|
| 510583858 | 2.4141 | 24.56±0.05 | 0.50±0.09 | 1.69 | $4.31^{+0.75}_{-2.90}$ | $18.6^{+2.2}_{-13.7}$ | 11.5±4.5 | $0.15^{+0.00}_{-0.05}$ | -2.40±0.16 | $316^{+478}_{-65}$ |
| 510838687 | 2.5539 | 24.69±0.05 | 1.45±0.09 | 2.58 | $0.78^{+0.35}_{-0.33}$ | $13.3^{+14.3}_{-5.6}$ | 12.5±6.0 | $0.10^{+0.05}_{-0.04}$ | -2.39±0.21 | $63^{+37}_{-45}$ |
| 511267982 | 2.8256 | 25.21±0.08 | 0.56±0.04 | 1.45 | $0.36^{+0.54}_{-0.13}$ | $15.4^{+4.2}_{-8.2}$ | 21.5±10.9 | $0.10^{+0.05}_{-0.04}$ | -1.62±0.29 | $25^{+101}_{-11}$ |
| 5100534435 | 2.9635 | 25.14±0.07 | 0.45±0.03 | 2.09 | $0.52^{+0.18}_{-0.19}$ | $17.9^{+3.1}_{-1.9}$ | 17.6±9.7 | $0.15^{+0.00}_{-0.00}$ | -1.90±0.29 | $32^{+18}_{-22}$ |
| 5100565880 | 3.0505 | 24.80±0.06 | <0.25 | 0.48 | $2.66^{+0.64}_{-0.96}$ | $8.1^{+3.9}_{-3.2}$ | 12.5±4.6 | $0.03^{+0.03}_{-0.03}$ | -2.40±0.20 | $501^{+293}_{-375}$ |
| 5100750978 | 2.9630 | 24.99±0.06 | 0.63±0.04 | 1.92 | $0.29^{+0.28}_{-0.08}$ | $9.7^{+7.4}_{-3.6}$ | 10.9±5.0 | $0.06^{+0.04}_{-0.03}$ | -2.50±0.18 | $32^{+47}_{-16}$ |
| 5100994378 | 2.7970 | 24.02±0.04 | 0.35±0.01 | 1.24 | $0.88^{+0.72}_{-0.25}$ | $60.2^{+11.8}_{-34.9}$ | 42.4±21.4 | $0.10^{+0.05}_{-0.00}$ | -1.94±0.13 | $16^{+47}_{-6}$ |
| 5100998761 | 2.4460 | 25.40±0.09 | <0.16 | 0.88 | $0.08^{+0.04}_{-0.02}$ | $7.8^{+0.2}_{-4.7}$ | 4.1±1.0 | $0.06^{+0.03}_{-0.00}$ | -2.77±0.29 | $10^{+5}_{-0}$ |
| 5101421970 | 2.4650 | 24.43±0.05 | 0.62±0.01 | 1.62 | $5.60^{+1.52}_{-1.33}$ | $15.0^{+3.0}_{-5.3}$ | 17.5±6.6 | $0.10^{+0.05}_{-0.05}$ | -2.22±0.15 | $159^{+635}_{-127}$ |
| 5101444192 | 3.4240 | 24.55±0.05 | <0.21 | 1.53 | $0.58^{+0.47}_{-0.04}$ | $61.2^{+4.6}_{-40.1}$ | 21.1±9.9 | $0.15^{+0.00}_{-0.05}$ | -2.32±0.33 | $10^{+3}_{-0}$ |

**Supplementary Table 1 | Main physical properties.** Columns show: (1) Identification number, (2) spectroscopic redshift, (3) F814W band AB magnitude, (4) circularized effective radius, (5) total radius, (6) stellar mass, (7-8) star formation rate, (9) stellar reddening, (10) UV slope, (11) age. The derivation of each of these quantities is presented in Methods.

| VUDS ID (1) | $EW_0$(Ly$\alpha$) (2) | $F$(Ly$\alpha$) (3) | $F$(CIV) (4) | $F$(He II) (5) | $F$(O III]) (6) | $F$(C III]) (7) | $12 + \log(O/H)$ (8) | $\log(C/O)$ (9) | $\log(U)$ (10) |
|---|---|---|---|---|---|---|---|---|---|
| 510583858 | 111±16 | 126.3±6.8 | 4.1±1.4 | 10.2±5.1 | 7.2±1.4 | 8.2±1.4 | 7.43±0.31 | -0.95±0.10 | -2.03±0.26 |
| 510838687 | 35±12 | 19.5±3.2 | 5.7±1.1 | <1.4 | 4.6±1.1 | 12.4±1.8 | 7.56±0.36 | -0.57±0.06 | -1.98±0.41 |
| 511267982 | 56±12 | 67.7±4.2 | 4.6±1.1 | <3.8 | 6.9±1.5 | 16.6±3.8 | 7.59±0.24 | -0.67±0.08 | -2.05±0.18 |
| 5100534435 | 47±11 | 44.0±4.0 | <1.4 | <2.0 | 4.0±0.8 | 4.6±1.9 | 7.71±0.20 | -1.00±0.09 | -2.03±0.13 |
| 5100565880 | 168±10 | 567.1±8.1 | 41.3±5.3 | 9.3±1.2 | 13.0±2.0 | 36.9±6.5 | 7.59±0.26 | -0.38±0.10 | -1.73±0.40 |
| 5100750978 | 82±12 | 120.9±5.9 | 1.6±0.8 | 2.0±1.2 | 4.0±1.2 | 7.9±1.2 | 7.55±0.21 | -0.78±0.04 | -2.25±0.11 |
| 5100994378 | 89±10 | 377.0±11.7 | 6.1±1.1 | 10.6±1.9 | 7.6±1.1 | 22.0±1.5 | 7.51±0.32 | -0.61±0.06 | -2.15±0.18 |
| 5100998761 | 185±35 | 361.5±21.4 | 8.3±1.4 | 6.9±3.1 | 8.6±1.7 | 21.4±2.4 | 7.38±0.19 | -0.64±0.06 | -2.01±0.11 |
| 5101421970 | 261±30 | 738.4±31.5 | 27.7±4.9 | 16.6±3.5 | 22.5±3.5 | 26.0±5.2 | 7.58±0.38 | -0.81±0.11 | -1.97±0.45 |
| 5101444192 | 267±35 | 182.3±3.5 | 3.5±0.9 | 2.7±0.4 | 4.9±0.9 | 11.5±1.3 | 7.66±0.24 | -0.67±0.07 | -2.04±0.16 |
| Composite | 131±11 | 260.0±2.0 | 9.1±1.1 | 7.0±1.2 | 8.2±1.0 | 15.2±1.2 | 7.54±0.25 | -0.69±0.08 | -1.90±0.27 |

**Supplementary Table 2 | Emission line measurements and derived abundances.** Columns show: (1) Identification number, (2) rest-frame equivalent width of the Lyman-$\alpha$ emission line, (3-7) observed emission line fluxes, (8) gas-phase metallicity, (9) carbon-to-oxygen ratio, (10) ionization parameter. The derivation of each of these quantities is presented in Methods.